\pdfoutput=1
\documentclass[prl,twocolumn,amsmath,amssymb]{revtex4-2}

\usepackage[normalem]{ulem} 
\usepackage{dcolumn}
\usepackage{graphicx}
\usepackage{bm}
\usepackage{mathtools}
\usepackage{amsmath,amsfonts}
\usepackage{tensor}
\usepackage{physics}
\usepackage{outlines}
\usepackage{natbib}
\usepackage{booktabs}
\usepackage{hyperref}
\usepackage[dvipsnames]{xcolor}
\usepackage{comment}
\usepackage{float}

\newcommand{\vct}[1] {\boldsymbol{#1}}

\usepackage[colorinlistoftodos]{todonotes}
\hypersetup{
    colorlinks=true,    
    linkcolor=blue,     
    citecolor=blue,     
    filecolor=magenta,  
    urlcolor=blue       
}
\begin{document}


\title{\small Factors Controlling the Statistics of Magnetic Reconnection in MHD Turbulence}

\author{M. B. Khan\textsuperscript{1}, 
M. A. Shay\textsuperscript{1},  
S. Oughton\textsuperscript{2}, 
W. H. Matthaeus\textsuperscript{1}, 
C. C. Haggerty\textsuperscript{3}, \\ S. Adhikari\textsuperscript{1}, P. A. 
Cassak\textsuperscript{4}, S.
Fordin\textsuperscript{1,5}, D. O$'$Donnell\textsuperscript{1}, Y. Yang\textsuperscript{1}, R. Bandyopadhyay\textsuperscript{1,6}, S. Roy\textsuperscript{1}}
\affiliation{\textsuperscript{1}Department of Physics and Astronomy, University
of Delaware, Newark, DE 19716-2570, USA}
\affiliation{\textsuperscript{2}Department of Mathematics, University of Waikato, Hamilton, New Zealand}
\affiliation{\textsuperscript{3}Institute for Astronomy, University of Hawaii at Manoa, Honolulu, HI 96822, USA}
\affiliation{\textsuperscript{4}Department of Physics and Astronomy and the Center for KINETIC Plasma Physics, Morgantown, WV, 26506, USA}
\affiliation{\textsuperscript{5}NASA Goddard Space Flight Center, Greenbelt, MD 20771, USA} 
\affiliation{\textsuperscript{6}Department of Astrophysical Sciences, Princeton University, Princeton NJ 08544, USA}

\begin{abstract}

\noindent 
We study the statistics of dynamical quantities associated with magnetic reconnection events embedded in a sea of strong background magnetohydrodynamic (MHD) turbulence using direct numerical simulations. We focus on the relationship of the reconnection properties to the statistics of global turbulent fields. For the first time, we show that the distribution in turbulence of reconnection rates (determined by upstream fields) is strongly correlated with the magnitude of the global turbulent magnetic field at the \emph{correlation} scale. The average reconnection rates, and associated dissipation rates, during turbulence are thus much larger than predicted by using turbulent magnetic field fluctuation amplitudes at the dissipation or kinetic scales. Magnetic reconnection may therefore be playing a major role in energy dissipation in astrophysical and heliospheric turbulence. 
 \end{abstract}

\maketitle
\emph{Introduction.} \label{sec:intro}
Magnetic reconnection
is a process by which 
stored magnetic energy is 
dynamically released as kinetic
energy, which may be either 
flow energy or thermal energy \citep{Parker-cmf,Biskamp1994MagneticReconnection}. Reconnection facilitates charged particle energization
and changes in magnetic topology,
and is thus considered an essential process in many heliospheric and astrophysical settings.
At the same time, the complex 
dynamical processes known as turbulence 
\citep{Frischbook, TuMarsch95}
occur
commonly in the same settings.
It is 
an appealing question to understand the 
relation between these fundamental 
features of cosmic electrodynamics. 

The problem is often posed in complementary ways:
Does turbulence influence the reconnection process?
And, what are the properties of reconnection occurring 
as an element of turbulence? 
There are several approaches 
that have addressed these
questions
\citep{MattLamkin86, Strauss88, LazarianVishniac99, ServidioEA09, LoureiroEA09, DaughtonEA11,KarimabadiEA14,  Zhdankin2013StatisticalTurbulence, LoureiroEA17, Boldyrev2017ApJ_844_125B}. These include 
adding low turbulence levels to standard reconnection geometry, 
imposing external driving in the reconnection region \cite{LoureiroEA09, sun2022physical} and 
evaluating mode-mode couplings among tearing modes. 
A different approach is to examine 
turbulence in a magnetofluid or plasma, and then identify and analyze reconnection activity embedded in the medium~(e.g., \citep{ServidioEA09}).

Here we adopt the 
latter approach 
wherein 
reconnection occurs
within a sea of interacting magnetic flux 
tubes \citep{ServidioEA10-recon, Wan2013GenerationTurbulence, Zhdankin2013StatisticalTurbulence, MalletEA17, LoureiroEA17, Dong2022Reconnection-drivenTurbulence}, a standard scenario in a large anisotropic 
turbulent plasma. We emphasize that in this case (and this study), the turbulence is dynamically determining the large MHD scale geometry which ultimately drives reconnection. This kind of study has been coined, ``reconnection as an element of turbulence.'' A fundamental question
is: what controls the rate of reconnection and its associated   energy release in such a system? \citep{ServidioEA10-recon}. We emphasize that the reconnection in such turbulence is local in nature, and may not be leading to a major change of magnetic connectivity at the global scales (e.g. dayside magnetosphere reconnection(e.g., \citep{Hesse2020MagneticFuture, LuoB2017}), and heliospheric current sheet reconnection (e.g., \citep{RaouafiEA23-parker}).
In this study, a two-dimensional (2D) magnetohydrodynamic (MHD) model of decaying turbulence is simulated. 
Reconnection X-points are diagnosed in the simulation, and their properties including reconnection rate are compared to the statistical properties of turbulence. The central finding,
not anticipated in many standard 
treatments (e.g., \citep{ShayEA18}), is that the reconnection rates are controlled by the 
dynamics of the large magnetic flux tubes at the 
correlation scale of the turbulence. This result implies that the dissipation of turbulence due to magnetic reconnection may be orders of magnitudes larger than previously thought.

\emph{Simulations.}\label{sec:setup}
We study the time evolution of a decaying turbulent system using 2D incompressible MHD simulations. For simplicity, we write the time evolution equations in terms of magnetic potential $a(x,y)$ and vorticity $\omega(x,y)$ with uniform density ($\rho=1$) as follows~\cite{Biskamp2003MagnetohydrodynamicTurbulence}:
\begin{align} \label{eq:VorticityEq}
\frac{\partial \omega}{\partial t}
 &= 
  - ( \vct{v}\cdot \vct{\nabla})\omega
  +
   ( \vct{b}\cdot \vct{\nabla})j
  +
   R_\nu^{-1}  \vct{\nabla}^2 \omega,
\\
 \label{eq:PotentialEq}
\frac{\partial a}{\partial t}
 &= -( \vct{v}\cdot \vct{\nabla})a
    + R_\mu^{-1}  \vct{\nabla}^2 a.
\end{align}
Here the magnetic field $ \vct{b}= \vct{\nabla}a\times \hat{\vct{z}}$, the velocity  $ \vct{v}= \vct{\nabla}\phi\times \hat{\vct{z}}$, the current density $j=- \vct{\nabla}^2 a$, and the stream function $\phi (x,y)$ is related to vorticity as $\omega=- \vct{\nabla}^2 \phi$. The system of equations is written in Alfvén units. The length scales are normalized to a characteristic length scale of the system, $L_0$. The velocity and magnetic fields are normalized to the root mean squared Alfvén speed $C_A$ and the time is normalized to $ L_0/C_A$. 
$R_\nu$ and $R_\mu$ are fluid and magnetic Reynolds numbers, respectively, and are also (non-dimensionalized) reciprocals of kinematic viscosity and resistivity. Note that any out-of-plane $(\hat{\vct{z}})$ magnetic field drops out of the dynamical equations for 2D incompressible MHD.

We solve these equations in a 2D periodic box of size $2\pi \times 2 \pi$, with equal values of magnetic and fluid Reynolds numbers $R_{\mu}=R_{\nu}=5000$, using a strongly energy-conserving pseudo-spectral code \cite{WanEA09-nongauss}, de-aliased using the 2/3 rule,  on a real-space grid with resolution $8192\times 8192$. In code units, the wavenumbers range from 1 to $k_\text{max}=(2/3)\times4096$. 
We 
ensure that the inequality 
  $ k_\text{max}/k_\text{diss}(t) \ge 3$, 
   is maintained,
  where $k_{\text{diss}} = R_{\nu}^{\frac{1}{2}} \langle \omega^2 + j^2 \rangle^{1/4}$ is the Kolmogorov dissipation wavenumber, 
  and $\langle \cdots \rangle$ denotes averaging over the simulation domain;
  this condition provides 
  accurate representation of statistics up to at least fourth order \cite{WanEA10-accuracy}.
  
  Initially, the energy is concentrated in a thin shell of $k$-radius $5 \leq k \leq 30$ and is equipartitioned between the fluid flow and magnetic fields. The system is time-advanced using a second-order Runge--Kutta scheme, and double precision is employed. 
We analyze the dynamics of our turbulent system at the time ($t = 0.3$) when the mean squared current density is near its maximum value. This is the point at which turbulence is well-developed, and energy is distributed across decades of scales as a near-power law. 
After this time, 
the global dynamics may be described approximately
by an adaptation of the von K\'arm\'an
similarity decay principle \citep{BandyopadhyayEA18-prx}.

A large number of magnetic reconnection events emerge from interactions between
large-scale magnetic structures (magnetic islands). Using techniques described in the Appendix, we find 625 X-points in the simulation, determine local the inflow
and outflow direction of each X-point, record the physical quantities associated with each
local reconnection geometry, and analyze their distribution within the context of the
turbulence pervading the entire system. In particular, we are interested in the statistics of upstream reconnecting fields and their relationship with the statistics of global turbulent magnetic fields. (Three X-points have zero reconnection rate and are excluded from the analysis.)

\begin{figure}
    \centering
    \hspace{0cm}
    \includegraphics[scale=0.63]{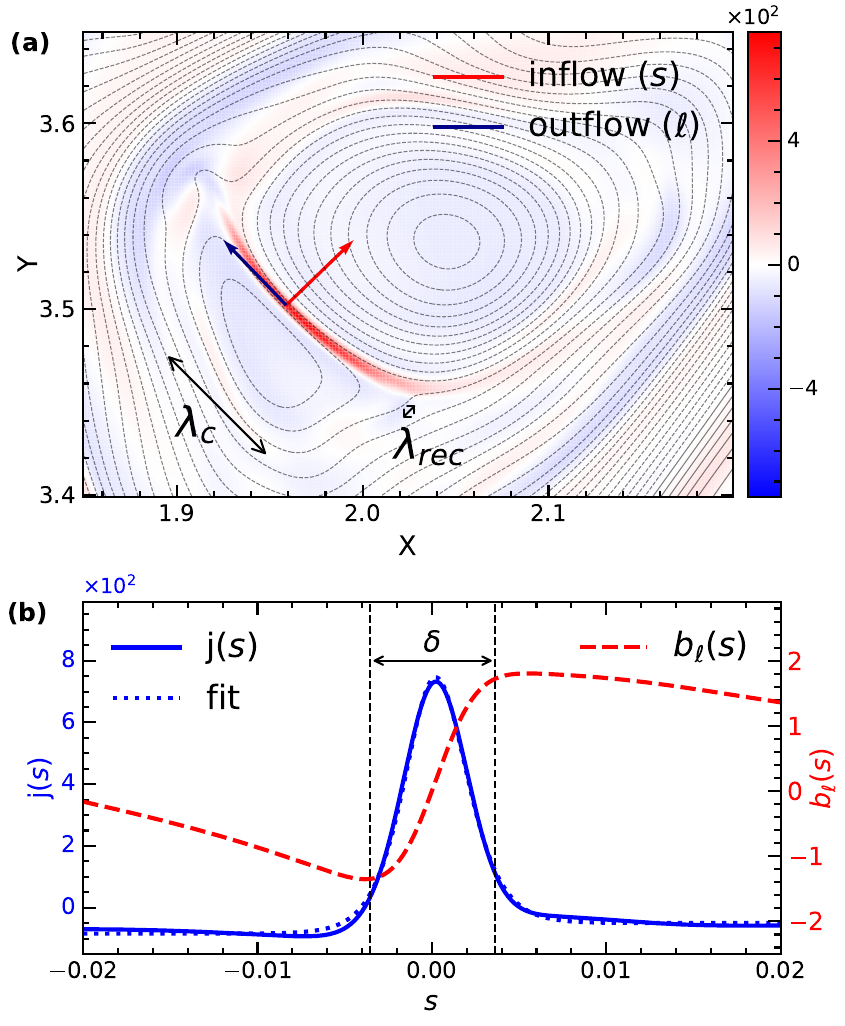}
    \caption{(a) Current density $j$ with contours of magnetic vector potential. $\lambda_c$ is the correlation length and $\lambda_\text{rec}$ is the average reconnection diffusion region thickness. Red and black arrows denote inflow and outflow directions associated with the diffusion region. 
    (b) One-dimensional cuts of current density $j$ (solid) and reconnecting magnetic field $b_\ell$ (dashed) along the inflow direction, 
    for the current sheet shown in panel (a),
    along with best fit line for $j(s)$ (dotted). Locations where $b_\text{up}$ is determined are shown by vertical dashed lines. }\label{fig:Current}

\end{figure}

\begin{figure}
    \centering
    \hspace{-1cm}
    \includegraphics[scale=0.75]{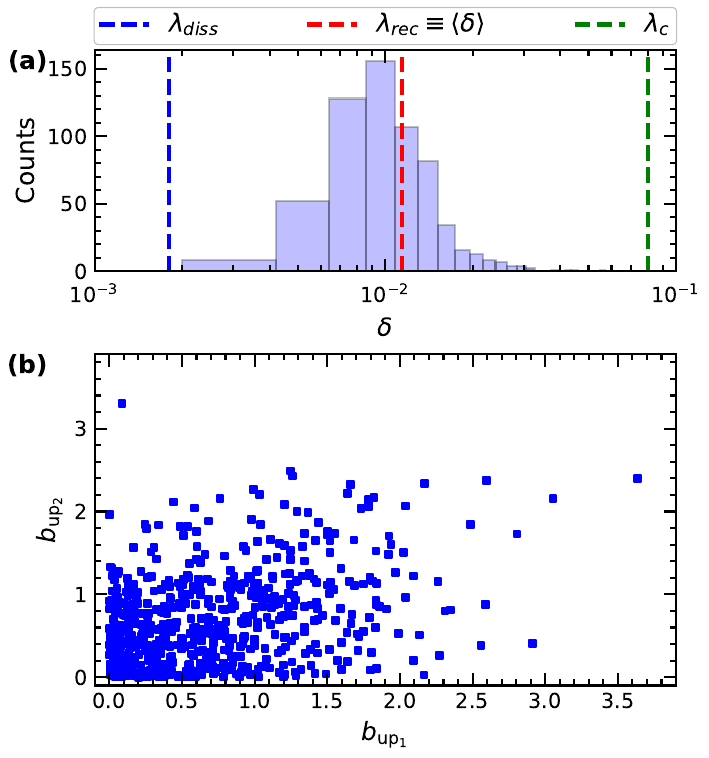}
    \caption{(a) Distribution of thicknesses of diffusion regions, with vertical lines showing the average thickness $\lambda_\text{rec}$, the dissipation scale $\lambda_\text{diss}$ and the correlation scale $\lambda_\text{c}$.
    (b) Scatter plot of $b_{\text{up}_1}$ vs $b_{\text{up}_2}$ 
    sampled at each reconnection site. 
    }
    \label{fig:tower}
\end{figure}

\begin{figure}
    \centering
    \hspace{-.4cm}
    \includegraphics[scale=0.61]{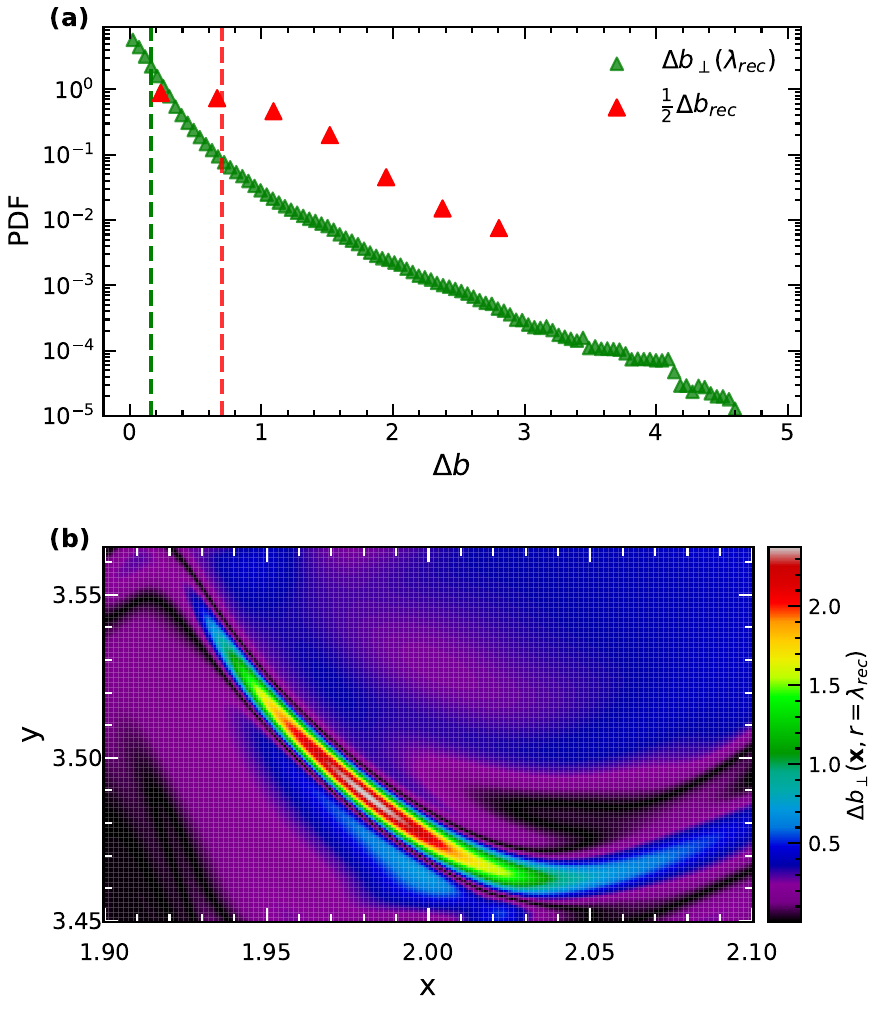}
    \caption{(a) Comparison of turbulence increments and reconnection magnetic fields. PDF of transverse turbulent magnetic field increments taken with lag $\lambda_\mathrm{rec}$ sampled over entire simulation domain, $\Delta b_\perp( \vct{x}, \lambda_\mathrm{rec})$. The reconnection magnetic field is defined as $\Delta b_\mathrm{rec}/2$ for each X-point. Average values of each quantity are shown with vertical dashed lines with $\Delta b_\perp( \vct{x}, \lambda_\mathrm{rec})$ averaged over $\vct{x}$ and $\Delta b_\mathrm{rec}/2$ averaged over X-points. 
     (b) 
     Turbulent field increments at transverse reconnection scale, $\Delta b_\perp( \vct{x}, \lambda_\text{rec}) $, near the strongly reconnecting current sheet shown in Fig.~\ref{fig:Current}a.}
     
      \label{fig:dbhist}
\end{figure}

\emph{Results.} \label{sec:results}
Fig.~\ref{fig:Current}a
illustrates an interaction between two large-scale magnetic structures, i.e., magnetic islands, during turbulence through the process of magnetic reconnection. A magnetic island is a coherent multi-scale structure within turbulence, encompassing a hierarchy of smaller-scale features. The typical size of a magnetic island in a turbulent system scales roughly with the correlation or energy-containing scale associated with the macroscopic system. The process of magnetic reconnection between two coherent magnetic structures, which might themselves be initially well decorrelated from one another, typically begins when their distance of closest approach nears the dissipation scale, $\lambda_\mathrm{diss}\equiv k_{\mathrm{diss}}^{-1}$. 
The current sheet thickness, i.e., the scale transverse to the reconnecting field, is usually found to be close to but somewhat larger than the dissipation scale of the system \citep{ServidioEA10-recon}.
The length of the 
reconnection regions is 
larger and 
broadly distributed 
around the correlation scale $\lambda_c$ \cite{ServidioEA10-recon}.

The relevance of both correlation scale and the inner scales of turbulence in the local reconnection process has been 
described previously \citep{ServidioEA10-recon}, but the quantitative implications for 
reconnection rates have not been addressed until the present study as far as we are aware.

To reveal these connections,
we identify the properties of all reconnection
sites, as described in detail in the Appendix.  For orientation, we refer to Fig.~\ref{fig:Current}a.
The positive directions of magnetic flux inflow ($s$) and outflow ($\ell$) at the reconnection site are marked by red and 
blue 
vector arrows, respectively. The correlation length $\lambda_c$ and average current sheet thickness $\lambda_\text{rec}$, i.e., the average transverse scale
associated with reconnection, are also shown for scale comparison.
Fig.~\ref{fig:Current}b 
indicates the diffusion region 
boundaries (vertical lines), superimposed on magnetic field and current profiles along the inflow direction for the reconnection site in Fig.~\ref{fig:Current}a. The upstream field strengths, $ b_{\text{up}_{1}} $ and $ b_{\text{up}_{2}} $, are evaluated at these boundaries, which are determined using the techniques outlined in the Appendix.
The profiles are typical of what is seen in a laminar magnetic reconnection setting,
 even though the system is turbulent (i.e., far from being a laminar flow).

Fig.~\ref{fig:tower}a shows the histogram of the diffusion region thickness, i.e., the transverse scale $\delta$ associated with the reconnection sites, computed using the curve-fitting technique. The histogram displays a relatively narrow distribution, centered around a mean value of  
$\lambda_\text{rec} \approx 0.01 \pm 0.001.$ Error estimation of $\lambda_\text{rec}$ is described in the Appendix.
Consequently,  $\lambda_\text{rec} \equiv \langle \delta \rangle$ provides a reliable estimate of the diffusion region thickness. The average transverse scale is a little more than 
5 times 
the dissipation scale, indicating that reconnection typically initiates well above the dissipation scale
(see also \cite{ServidioEA10-recon}).

A scatter plot of upstream reconnecting fields $b_{\text{up}_1}$ and $b_{\text{up}_2}$ sampled at each reconnection site is  shown in Fig.~\ref{fig:tower}b. 
There is no significant  correlation between the two upstream reconnecting fields, even though the upstream fields are separated by a distance much less than the correlation scale.
An explanation is that 
these fields originate in 
two large, distinct, coherent island
structures that are not strongly related to one another.  
At this
point the hypothesis 
emerges that 
the global turbulent field statistics at the scale of 
reconnecting current sheets
thicknesses may not be a good estimator of the 
reconnecting fields, and therefore 
will not provide an accurate basis for estimating reconnection rates.

To pursue this line of reasoning
requires estimation
of the correlation scale, by first computing the autocorrelation function 
$ C(r) =  \langle \vct{b}(\vct{x}+\vct{r}) \cdot  \vct{b}(\vct{x}) \rangle/\langle \vct{b}(\vct{x})\cdot\vct{b}(\vct{x}) \rangle$, where $\vct{r}$ is the lag vector. The correlation scale is then obtained by integrating the autocorrelation function $\lambda_c = \int_0^\pi C(r)\, \mathrm{d}r$, where $\pi$ is half the size of the periodic box; for the simulation time studied, $\lambda_c \approx 0.08$. 
Since the turbulence is not biased 
to any position or direction, it may be viewed as homogeneous and isotropic.
Therefore the direction of lag vector $\vct{r}$ is not important in the computation of the 
correlation function, or of the global turbulence increments (introduced in what follows). 

The next step is to explore the statistical relationship between the upstream reconnecting fields and the global turbulent fields. To achieve this,
we 
compare increments of magnetic field components
computed in two distinct ways:
One, to make contact with 
standard theories of reconnection, 
we compute the \emph{reconnection increments}, defined 
as transverse magnetic field increments 
across the reconnecting current sheets, employing the average upstream field magnitude defined briefly in the next paragraph and in detail in the Appendix. Two, to characterize the turbulence we compute the turbulence increments i.e., the increments of \emph{global} turbulent magnetic fields over the entire simulation, as defined below. 

One, The reconnection increments 
    $ \Delta {b}_{\text{rec}}$, computed for each X-point,  
    are calculated by taking the difference of
upstream magnetic field evaluated at the boundaries of diffusion region as 
 $  \Delta b_\text{rec}  = | \vct{b}_\mathrm{up_2}  -  \vct{b}_\mathrm{up_1} | $ 
 or 
 $  \Delta b_\text{rec}  = | \vct{b}_\mathrm{up_2} | + | \vct{b}_\mathrm{up_1} | $. 
The average upstream magnetic field at each X-point is then given by $\frac {1}{2} \Delta {b}_{\text{rec}}$. Note that at each reconnection site, the  lag vector is aligned with the local inflow direction, i.e., perpendicular to the reconnecting magnetic field. This choice makes the reconnection increments inherently transverse. 

Two, 
the \emph{global turbulence increments} are defined as 
$  \Delta \vct{b}( \vct{x}, \vct{r}) = 
    \vct{b}(\vct{x} + \vct{r}) - \vct{b}(\vct{x})$. We define the \emph{transverse increment} 
  $ \Delta {b}_\perp( \vct{x},\vct{r} )$ and the longitudinal increment $ \Delta {b}_\parallel( \vct{x},\vct{r} )$ as the projection of 
 $ \Delta \vct{b}(\vct{x}, \vct{r} )$ 
 perpendicular and parallel to the lag vector $\vct{r}$, respectively. Note that different from reconnection increments, in this case the lag vector $\vct{r}$ is a constant across the whole simulation. For simplicity and
 consistency with the reconnection increments,
 we will examine the absolute value of the global increments
 and discard the ``$\mid\cdots \mid$.'' 
 For the present purposes the transverse turbulence increments are examined 
at two different lags: the average transverse reconnection scale $\lambda_\text{rec}$ and the correlation scale $\lambda_c$.

We now proceed to compare statistics of the
 turbulence increments (at two scales) with statistics of the reconnection increments. Note that the transverse turbulence increments are effectively sampling only a single component of the turbulent magnetic field; therefore we compare this global increment with the average upstream reconnection field $\Delta b_\textrm{rec}/2$.

The red triangles in Fig.~\ref{fig:dbhist}a are the probability density function (PDF) of $\Delta b_\text{rec}/2$ for all reconnection sites. The broad distribution reflects the wide variability of upstream reconnecting fields within the system. The distribution also exhibits a strong tail, suggesting the presence of significant intermittent reconnection events. These tail values indicate that while most field increments are moderate, there are occasional instances of particularly intense reconnection, contributing to the overall variability observed. 

In Fig.~\ref{fig:dbhist}a, this PDF of $\Delta b_\text{rec}/2$ is compared with the PDF of the transverse increments of turbulent fields at the average current sheet thickness $\lambda_\text{rec}$, computed globally. The mean values of the PDFs are indicated by vertical lines. Clearly, the turbulent magnetic fields have a very different statistical distribution than the reconnection fields. Both the shape and average values of the distributions do not match.

We emphasize that
it might seem reasonable to have postulated that 
the reconnecting fields 
correspond to 
the turbulence increments evaluated at 
the transverse reconnection scale,
  $ \lambda_\text{rec} $. 
However, the result in 
Fig.~\ref{fig:dbhist}a 
clearly  demonstrates that this assumption fails.
This mismatch between the turbulence and reconnection statistics occurs because at small lags the neighborhood near a current sheet exhibits a wide range of turbulent increments depending on where it is sampled. In Fig.~\ref{fig:dbhist}b, the turbulent increments $\Delta {b}_\perp( \vct{x},\lambda_{\mathrm{rec}} )$ are plotted in a close vicinity of the strongly reconnecting field line in Fig.~\ref{fig:Current}a. The near-current sheet dynamics generate increments ranging in value from about 0.5 to 2.5. In contrast, this current sheet generates only a single reconnection increment near the peak value of that range. The PDF of turbulent magnetic field increments will thus be dominated by relatively small values, leading in Fig.~\ref{fig:dbhist}a to a distribution more strongly peaked near $\Delta b_\perp(\lambda_{\mathrm{rec}})=0$ and also a much smaller average value.

\begin{figure}
    \centering
    \hspace{-.4cm}
    \includegraphics[scale=0.56]{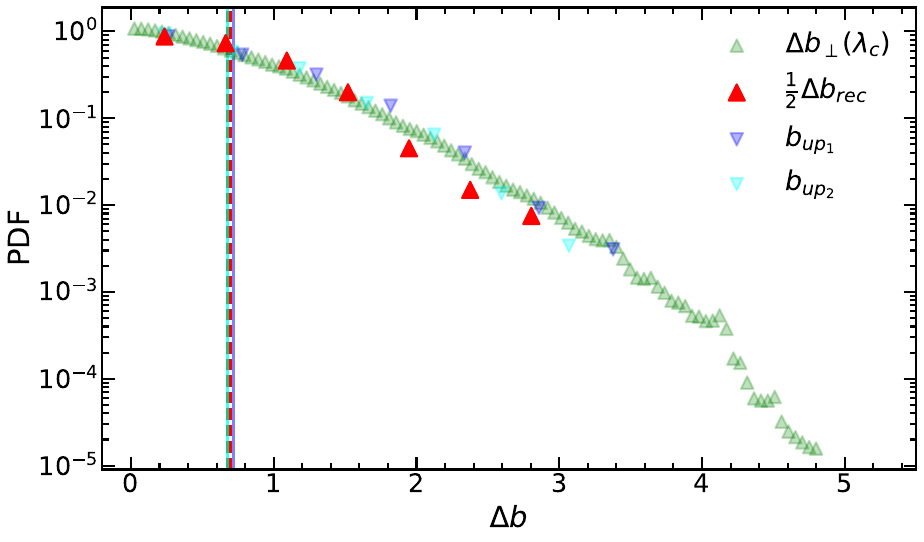}
    \caption{ PDFs of transverse turbulence increments at the correlation scale compared with PDFs of reconnection fields $\Delta b_\textrm{rec}/2,$ $b_1$, and $b_2$. Average values of each quantity are shown as vertical lines with matching color.  The agreement of these distributions is striking, and in significant contrast with Fig.~\ref{fig:dbhist}(a). }\label{fig:pdfs}
\end{figure}

The above analysis hints that the reconnection may be fundamentally linked to larger scale turbulent dynamics. This idea is examined in Fig.~\ref{fig:pdfs}, which shows PDFs of $\Delta b_\text{rec}/2$ and $\Delta b_{\perp} (\vct{x}, \lambda_c)$. Additionally, we plot the PDFs of the individual contributions to the upstream field increments at each reconnection site, namely $b_{\text{up}_1}$ and $b_{\text{up}_2}$. The result is striking. The reconnection magnetic fields on each side of the X-point as well as their average value exhibit PDFs that match both the shape and average value of the PDFs of the turbulent magnetic fields.

Clearly, the statistics of reconnection are not coupled to the turbulence properties evaluated at the transverse reconnection scale, but instead are directly controlled by the statistics of the turbulent magnetic field at the correlation scale. This is the central result of this paper.

\emph{Discussion.} 
The results of this study demonstrate that the driver of reconnection, here shown to be the large scale, or \emph{energy-containing} fluctuations, coincides with the driver of the turbulence cascade. To establish this connection, we  analyze the statistics of small-scale reconnection events in a decaying turbulence simulation.   Comparing the  PDFs of (global) increments at both the correlation scale and at the average transverse reconnection scale with the PDF of the reconnection upstream field gives the key result of the study: the PDF of fields associated with reconnection aligns closely with the PDF of turbulence increments at the correlation scale, but does not align well with turbulence increments at  the transverse reconnection scale. Additionally, the average value of the reconnection upstream fields matches well with the average value of turbulent fields at correlation scales, and is much greater than that at the transverse reconnection scale. This supports the conclusion that, for the present numerical experiment,  the energy-containing scale dynamics are regulating the statistics of the reconnection.  We suspect that this result may be applicable far beyond a 2D turbulent MHD system. 

These results have significant implications for turbulent astrophysical plasmas. During reconnection, the reconnection rate and resultant heating are strongly controlled by the upstream reconnection field. A Sweet--Parker analysis shows that the reconnection rate is proportional to $b_\text{up}^2$ (e.g., \cite{YamadaEA10}). The temperature change from the inflow region to the exhaust is also proportional to $b_\text{up}^2$ 
  \cite{PhanEA14, ShayEA14, HaggertyEA15}, while the heating rate of plasma due to reconnection is proportional to $b_\text{up}^3$ 
  \cite{ShayEA18,StawarzEA22}.
 A strong potential connection to classical 
 turbulence theory becomes evident. 
 The latter expression may be written, based on the findings above, in terms of r.m.s. fluctuation Alfv\'en speed $C_A$ 
 evaluated at the controlling length, here 
identified as the correlation scale $\lambda_c$. Then we can identify the  heating rate per unit mass 
 as $ \propto C_A^3/\lambda_c$. This is the magnetofluid form of the classical von K\'arm\'an turbulence heating rate \citep{KarmanHowarth38,HossainEA95}.

The upstream reconnection magnetic field is often estimated as the amplitude of the turbulent field at the transverse reconnection scale ($\lambda_\text{rec}$ in our notation) (e.g., \citep{ShayEA18}). Such a small scale leads to a very small reconnection field and thus small heating rates. However, the amplitude of the turbulent field at the correlation scale is significantly larger, with associated higher dissipation rates. To put this in perspective, we estimate the ratio of the turbulence amplitude in the solar wind for the correlation scale versus the transverse reconnection scale. Using a Kolmogorov type analysis, nominal solar wind conditions at 1\,AU, and estimating the reconnection scale as the ion inertial length $d_i$,

\begin{equation}
\label{eq:deltaB-ratio}
\frac{ b_{\lambda_{c}}}
     { b_{\lambda_\text{rec}}}
  \sim 
 \left( \frac{\lambda_c}
             {\lambda_{d_i}}\right)^{\frac{1}{3}}
  \sim 
\left( \frac{10^6\,\text{km}}
            {10^2\,\text{km}} \right)^\frac{1}{3}
 \sim 
   20 .
\end{equation}

If such a scaling is applicable, Eq.~\eqref{eq:deltaB-ratio} implies that the dissipation of turbulent fluctuations due to reconnection may be much larger than previous thought: reconnection rates and reconnection temperature changes may be about 500 times larger than previous estimates, while heating rates may be 10,000 times larger. Energy dissipation through reconnection therefore is likely to play a major role in dissipating the turbulence energy cascade.

\emph{Acknowledgments.}
This research is supported by NASA grants 80NSSC20K0198 (MAS), 80NSSC20K1813 (MAS), 80NSSC24K0559 (MAS), 80NSSC19K0284 (WHM), 80NSSC21K0739 (RB), and 80NSSC21K1458 (RB); NSF grants AGS-2024198 (MAS), AGS-2108834 (WHM), PHY-2205991 (CCH), AGS-1936393 (CCH), AGS-2338131 (CCH), PHY-2308669 (PAC); 80NSSC24K0172 (PAC and MAS), the 2024 Ralph E. Powe Award (YY); and the  University of Delaware General University Research Program Grant (YY). We acknowledge high-performance computing support from the Derecho system \href{https://doi.org/10.5065/qx9a-pg09}{(https://doi.org/10.5065/qx9a-pg09)} provided by the NSF National Center for Atmospheric Research (NCAR), sponsored by the National Science Foundation. We also acknowledge the National Energy Research Scientific Computing Center (NERSC), a U.S. Department of Energy Office of Science User Facility operated under Contract No.~DE-AC02-05CH11231.

\emph{Data Availability Statement.} The data that support the findings of this study are openly available in the Zenodo repository and can be accessed via Ref.~\cite{khan2025_data}.



 \newcommand{\BIBand} {and} 
%


\clearpage

\appendix

\section{Appendix A: Determining Reconnection Statistics}

\label{sec:apndx}
Here we provide details of the technique used to determine the magnitude of the reconnection magnetic fields at the upstream edges of the diffusion region associated with each X-point. Because the reconnection is often asymmetric, each X-point has two distinct upstream fields, $b_\mathrm{up_1}$ and $b_\mathrm{up_2}$. The values then give the reconnection increments $\Delta b_\mathrm{rec} = |b_\mathrm{up_2}| + |b_\mathrm{up_1}|$. 

\begin{figure}
    \centering
    \includegraphics[scale=0.49]{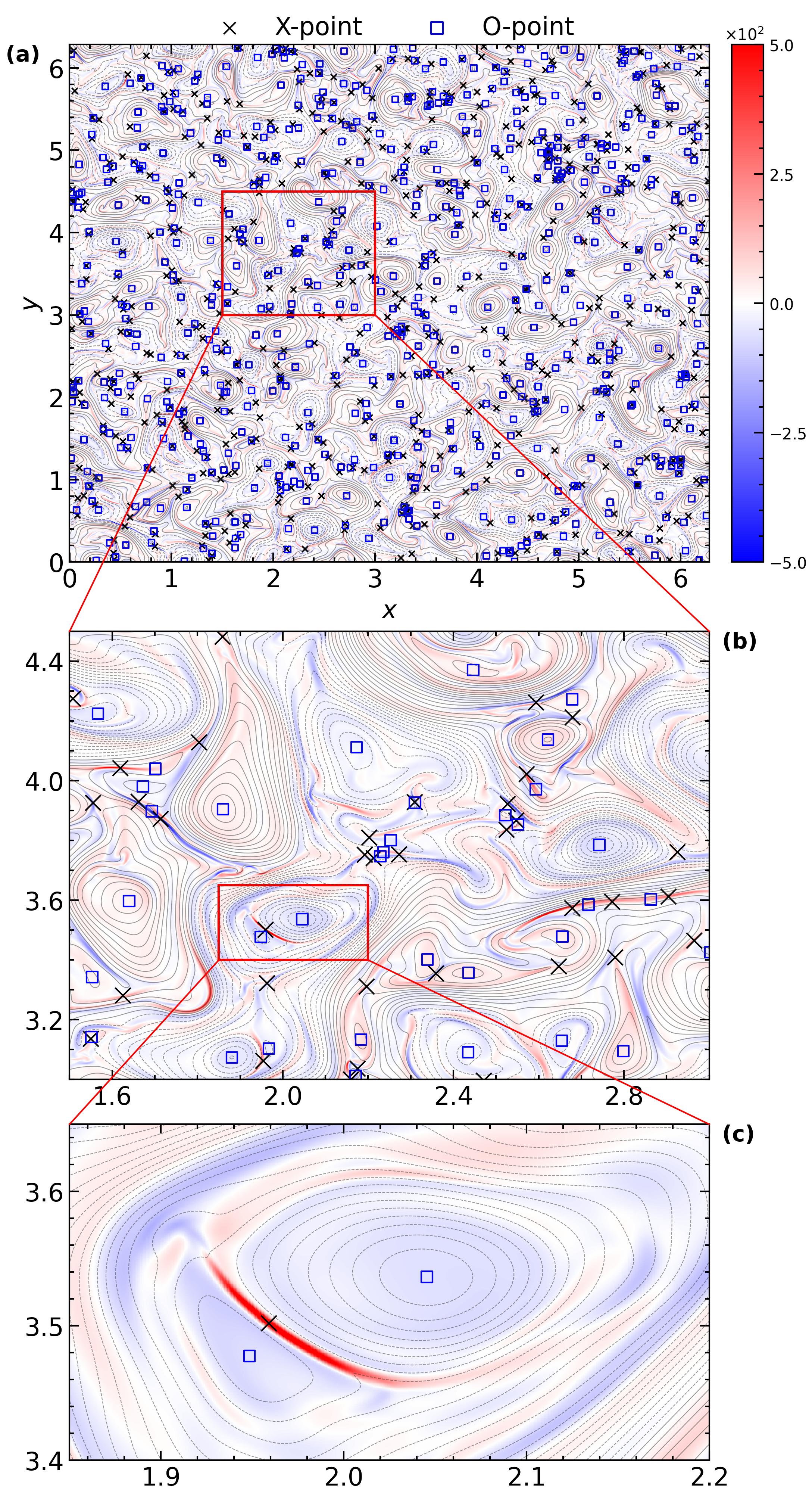} 
    \caption{Current density $j$ at the time of analysis ($t = 0.3$) with contours of the magnetic potential $a$ superimposed. The X-points and O-points are marked by `$\times$' and `$\square$' respectively. Insets highlight subsections of the system.}
    \label{fig:oview}
\end{figure}

First, the X-points are found by examining the extrema of the out-of-plane magnetic vector
potential $(\nabla a = 0)$, with X-points located at saddle
points~\cite{ServidioEA09,ServidioEA10-recon, HaggertyEA17}. Because X-points are usually
located between grid points, in order to determine their location with a higher accuracy,
an interpolation is used to double the number of grid points in each direction ($2N \times
2N$) by padding the Fourier transform of the vector potential with zeros. Note that interpolation to an even larger grid than ($2N \times2N$) does not yield any benefit; a test of $4N \times 4N$
interpolation revealed  little or no change to the number of X-points, $\lambda_\mathrm{rec}$, nor the primary conclusions. The ``sea'' of X-points and O-points which are found with this method are shown in Fig.~\ref{fig:oview}a overlaid on a color map of $j$ with contours of $a.$ To highlight
the large range of scales in the system, we show successive magnification of a small region of the total system
(Fig.~\ref{fig:oview}b and ~\ref{fig:oview}c). Ultimately, a single X-point is revealed
in the presence of the global turbulence.

Second, using the Hessian matrix at each X-point, the local inflow $(\hat{\vct{s}})$ and outflow $(\hat{\vct{\ell}})$ directions are determined, as described in detail in~\citep{ServidioEA10-recon}. An example of $(\hat{\vct{s}})$ and $(\hat{\vct{\ell}})$ for a single X-point is shown in Fig.~\ref{fig:Current}a. We emphasize that these vectors are different for every X-point.  

\begin{figure}
    \hspace{-0.2cm}
    \includegraphics[scale=0.45]{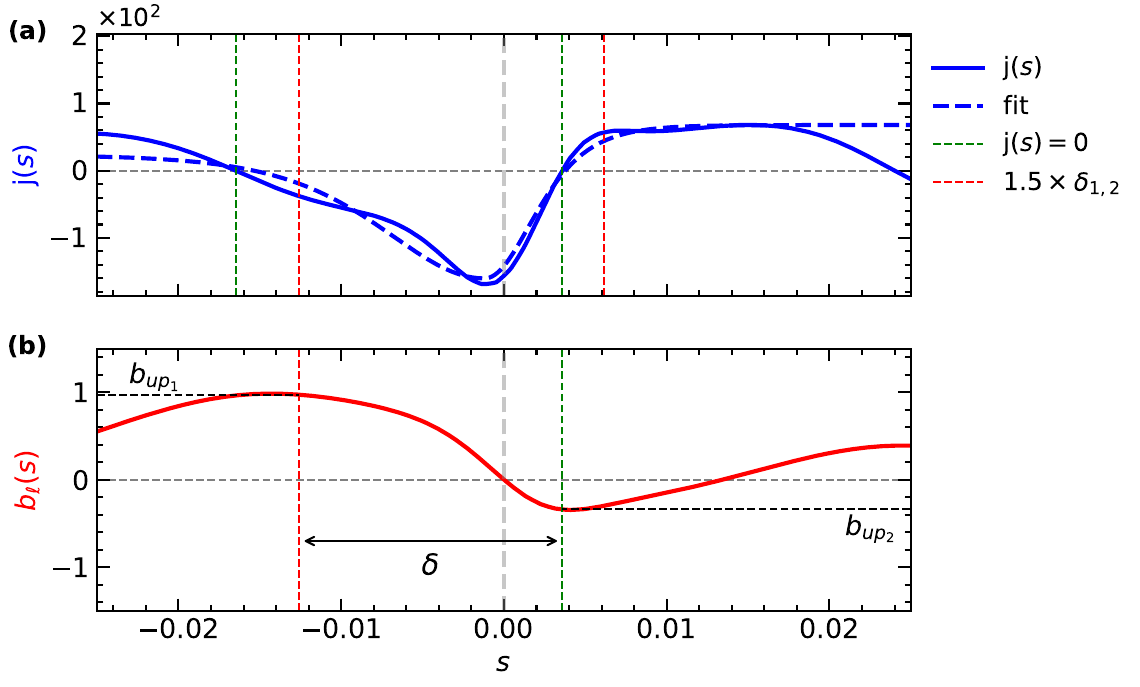} 

    \caption{Determination of upstream magnetic fields for a representative X-point. One dimensional cuts through the X-point along the inflow direction $(\hat{\vct{s}})$. X-point location denoted by vertical gray lines at center. (a) current density $j(s)$ together with the best fit curve. Vertical green and red lines denote possible choices for upstream edges of diffusion region. (b) Reconnecting magnetic field $b_\ell(s)$. The chosen diffusion region boundaries are denoted by the green and red vertical lines ($1.5\,\delta_1$ for $s < 0$ and $j = 0$ for $s >0$). Horizontal black lines indicate the resultant $b_\textrm{up1}$ and $b_\textrm{up2}$. }
    \label{fig:determineBup}
\end{figure}

While some X-points have symmetric and simple variation in the magnetic field, like the case in Fig.~\ref{fig:Current}b, other X-points can be much more complicated, an example of which is shown in Fig.~\ref{fig:determineBup}. In this strongly asymmetric case, the peak of current $j$ is slightly offset from the X-point and $j$ has a complex structure as one moves away from the X-point. For $s < 0$ the current continuously increases, ultimately becoming positive for $s < -0.016$. For $s > 0$ it quickly changes sign leading to a plateau of positive $j$ starting near $s = 0.01$.

There are two primary ways to determine the upstream edge of the diffusion region: (1) performing a fit to determine the approximate current width, or (2) taking the location where $j = 0.$ However, for a complex case such as in Fig.~\ref{fig:determineBup}a, we have found that a hybrid of the two methods is necessary to accurately classify complex X-points with relatively low reconnection rates. 

In this hybrid method, the current $j(s)$ is fitted to a continuous but piecewise function broken into a left and right half at the location where $j$ is maximum, denoted as $s = s_p$. 
\begin{equation}
f(s) =
\begin{cases}
  A_1 \sech^2 \left( \frac{s-s_{p}}{\delta_1} \right) - A_1 + C, & s<s_p \\
  A_2 \sech^2 \left( \frac{s-s_p}{\delta_2} \right) - A_2 + C, & s>s_p 
\end{cases}
\end{equation}
The peak current amplitude $C$, the function amplitudes $A_1$ and $A_2$, and the current sheet thicknesses $\delta_1$ and $\delta_2$ are all fitting parameters. The current profile is iteratively fitted to the function for $|s| < 0.01$.

The fitted function is shown as the dashed blue line in Fig.~\ref{fig:determineBup}a. The two criterion for the diffusion region edge are also shown as vertical dashed colored lines: criterion A (green) is  the location where $j = 0$ and Criterion B (red) is a distance $1.5\,\delta_1 $ or $1.5\,\delta_2$ as appropriate. Independently on each side of the X-point, the diffusion region edges are chosen using the criterion which is closest to the X-point. The resulting diffusion region width $\delta$ and the values of $b_\mathrm{up_1}$ and $b_\mathrm{up_2}$ can be determined from $b_\ell(s),$ as shown in Fig.~\ref{fig:determineBup}b. 

The error in $\delta$ is estimated as follows. For criterion A the error is simply associated with the grid scale $\Delta=\frac{2 \pi}{8192}$, so an error of $\Delta/2$ is used. For criterion B, the error is calculated from the covariance matrix of the curve fit. The uncertainty of the total diffusion width $\delta$ is then calculated through error propagation such that: 
 $ \sigma_\delta = \sqrt{ (c_1\sigma_{\delta_1})^2 
      + (c_2\sigma_{\delta_2})^2 
      + 2 c_1 c_2 \, \textrm{cov}( \delta_1, \delta_2 )}$ 
where the values of $c_{1,2}$ depend upon the criterion used to determine the boundary at either side: $c_{1,2}=1$ for criterion A and $c_{1,2}=1.5$ for criterion B. The result is a diffusion region thickness and its error for each X-point, which when averaged appropriately, leads to an average diffusion region thickness of $\lambda_\mathrm{rec} \equiv \langle \delta \rangle = 0.01 \pm 0.001$.

\begin{figure}[H]
    \centering
    \hspace{-.4cm}
    \includegraphics[scale=0.56]{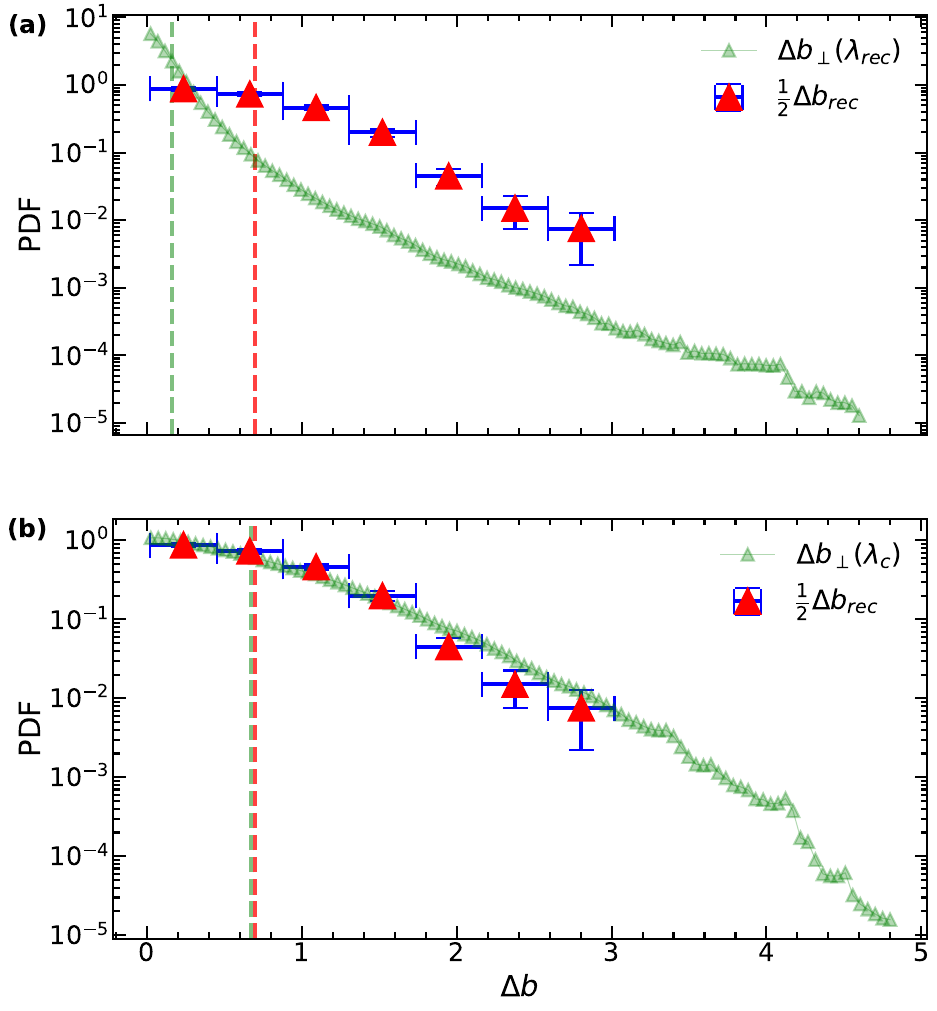}
    \caption{  PDFs of reconnection fields $\Delta b_\textrm{rec}/2$ with error bars compared with transverse turbulence increments with lags of (a) $\lambda_\textrm{rec}$ and (b) the correlation scale $\lambda_c$. Average values of each quantity are shown as vertical lines with matching color.  }
    \label{fig:error-bars}
\end{figure}

The errors associated with the probability distribution functions (PDFs) of the
reconnection increments in Figs.~\ref{fig:dbhist} and~\ref{fig:pdfs} can also be
estimated. The error in $\Delta b$ is 0.21, which is half of the bin size used for the PDF.   The
vertical error is due to counting statistics. Note that the vertical error is too small to
be visible  for $\Delta b_\textrm{rec} \lesssim 1$. The PDFs with error bars are shown in Fig.~\ref{fig:error-bars}.

\section{Appendix B: Investigating the effects of selection bias}
\begin{figure}[H]
    \centering
    \vspace{-.1cm}
    \hspace{-1cm}
    \includegraphics[scale=0.55]{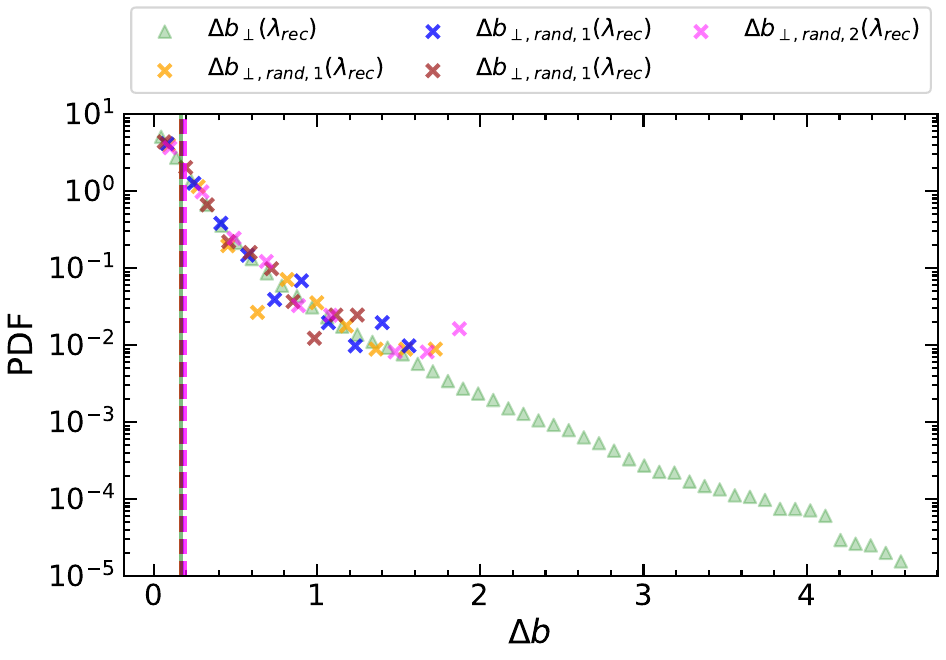}
    \caption{
    Probability density function (PDF) of transverse turbulence increments $\Delta b_\perp$ at the average transverse reconnection scale $\lambda_{\mathrm{rec}}$. The global PDF (green) is compared with three independent random subsets of 622 samples (orange, blue, brown; first approach) and with 622 transverse increments sampled at random locations within the global system (magenta; second approach). Average values are shown as vertical lines with corresponding colors. Note that the average values are so similar that the vertical lines are nearly indistinguishable from each other.}
    \label{fig:example}
\end{figure}

The central result of this paper is that the distribution of magnetic fields responsible for magnetic reconnection in a 2D MHD turbulent system correlates strongly with the global turbulent field at the \emph{energy containing} scales of the system, as opposed to the turbulent fields at the transverse reconnection scale.  
The statistical analysis of magnetic reconnection in this letter is based on a total of 622 events. However, the statistical analysis of the global field consists of a significantly larger number of samples compared to the ensemble of reconnection events. This raises the question of whether the observed correlation is a consequence of selection bias. To address this concern, we studied the effect on the PDF of turbulence increments due to random downsampling of the dataset to 622 increments, equal to the total number of reconnection events in our system. In all cases we examined a lag equal to the average transverse reconnection scale. 
We downsampled the data using two different methods. 
First, we randomly sampled 622 increments multiple times from the ensemble of global increments at the transverse reconnection scale. Second, we sampled the transverse increments at 622 randomly chosen locations and directions within the system, using a procedure analogous to that used for computing reconnection increments but without restricting the locations to X-points. In Fig~\ref{fig:example}, we then compared all of these randomly selected increments to the global probability density function (PDF) and found that they closely matched the global PDF. This analysis robustly eliminates the possibility of selection bias.

\hspace{1cm}

\end{document}